\newcommand{\ga}{\gamma}

\newcommand{\de}{\delta}

\newcommand{\la}{\lambda}

\newcommand{\om}{\omega}

\newcommand{\pa}{\partial}

\newcommand{\been}{\begin{equation}}
\newcommand{\een}{\end{equation}}
\newcommand{\beena}{\begin{eqnarray}}
\newcommand{\eena}{\end{eqnarray}}

\newcommand{\tn}{\textnormal}

\newcommand{\deriv}[2]{\frac{\partial{#1}}{\partial{#2}}}
\newcommand{\derivtwo}[2]{\frac{\partial^{2}{#1}}{\partial{#2}^2}}
\newcommand{\derivtwomix}[3]{\frac{\partial^{2}{#1}}{\partial{#2}\partial{#3}}}

\documentclass{article}

\usepackage{amsmath,amssymb}
\usepackage{verbatim}
\usepackage{graphicx}
\usepackage{psfrag}

\DeclareMathOperator{\Div}{div}
\DeclareMathOperator{\Grad}{Grad}
\DeclareMathOperator{\grad}{grad}

\author{W.J.~Parnell\\
\footnotesize{School of Mathematics, University of Manchester}\\\footnotesize{Oxford Road, Manchester, M13 9PL.}}
\title{Nonlinear pre-stress for cloaking from antiplane elastic waves}
\date{19th October 2011}

\begin{document}

%\label{firstpage}

\maketitle

\numberwithin{equation}{section}

\begin{abstract}
A theory is presented showing that cloaking of objects from antiplane elastic waves can be achieved by elastic pre-stress of a neo-Hookean nonlinear elastic material. This approach would appear to eliminate the requirement of metamaterials with inhomogeneous anisotropic shear moduli and density. Waves in the pre-stressed medium are bent around the cloaked region by inducing inhomogeneous stress fields via pre-stress. The equation governing antiplane waves in the pre-stressed medium is equivalent to the antiplane equation in an unstressed medium with inhomogeneous and anisotropic shear modulus and isotropic scalar mass density. Note however that these properties are induced naturally by the pre-stress. Since the magnitude of pre-stress can be altered at will, this enables objects of varying size and shape to be cloaked by placing them inside the fluid-filled deformed cavity region.
\end{abstract}


\section{Introduction}

In recent years there has been a great deal of interest in theoretical and practical issues associated with the cloaking of objects when subject to a variety of incident wave fields. The original theory was developed in the context of electromagnetic waves and used the coordinate transformation principle (Greenleaf et al.\ 2003, Leonhardt 2006, Pendry et al.\ 2006, Cummer et al.\ 2006). Attention then shifted onto the possibility of cloaking in acoustics (see e.g.\ Cummer and Schurig 2007, Chen and Chan 2007, Cummer et al.\ 2008, Torrent and Sanchez-Dehesa 2008, Norris 2008, 2009), surface waves in fluids (Farhat et al.\ 2008) and elastodynamics (Milton et al.\ 2006, Brun et al.\ 2009, Norris and Shuvalov 2011). It was shown by Milton et al.\ 2006 that the latter is more difficult to achieve than the other physical applications mentioned due to the lack of invariance of Navier's equations under general coordinate transformations, unlike the equations governing electromagnetic and acoustic waves. This explains the relative smaller number of studies in elastodynamic cloaking. A special case is that of flexural waves in thin plates (Farhat et al.\ 2009). Furthermore invariance of the governing equations \textit{can} be achieved for more a specific class of transformations if assumptions are relaxed on the minor symmetries of the elastic modulus tensor as was described for the in-plane problem by Brun et al.\ (2009). Norris and Shuvalov (2011) exploited this by using Cosserat materials.

It is well-known that cloaking can be achieved for antiplane elastic waves from a cylindrical region (using a cylindrical cloak) in two dimensions (Brun et al.\ 2009). This can immediately be recognized by the duality between antiplane waves and acoustics in this dimension. As has been indicated above and as will be shown in section \ref{sec:classical} of the present paper, in order to achieve cloaking by the method of coordinate transformations, the cloak must have an anisotropic, inhomogeneous shear modulus and scalar, inhomogeneous density. Indeed, inhomogeneity and anisotropy are standard requirements of cloaking materials. Furthermore the mappings that require this inhomogeneity are singular in the sense that some of these moduli are unbounded on the inner surface of the cloak (Greenleaf et al.\ 2003, Pendry et al.\ 2006, Cummer et al.\ 2006). In reality this condition must be relaxed so that they are merely required to be very large near the inner region of the cloak thus yielding non-perfect cloaking. Construction of the cloaking metamaterial itself is non-trivial, although progress has been made in this area in some special cases (see e.g.\ Schurig et al.\ 2006, Farhat et al.\ 2008, Zhang et al.\ 2011).

Here we present an alternative scheme to produce cloaking from antiplane elastic waves by the use of nonlinear elastic pre-stress. The nonlinear pre-stress generates an anisotropic and inhomogeneous (incremental) shear modulus but the density remains \textit{homogeneous}. Furthermore, these properties arise naturally as a result of the pre-stress and so we are not required to construct a special metamaterial cloak. Additionally the cloaking is \textit{active} in the sense that the size of the cloaked region can be altered by modifying the pressure inside this region.

We begin with a small cylindrical cavity of radius $A$ located at the origin of a soft incompressible nonlinear elastic medium  whose constitutive behaviour is described by a neo-Hookean strain energy function (Ogden 1997, 2007). With reference to figure \ref{fig2}, a line source is located at a distance $R_0$ from the origin and angle $\Theta_0$ subtended from the $X$ axis. The cavity radius $A$ is chosen so that scattering from the line source is weak. This can always be achieved by ensuring that $KA\ll 1$ where $K$ is the wavenumber of the soft rubbery medium. We then increase the pressure inside the cavity (by increasing internal fluid pressure for example) so that the radius of the cavity region increases to $a$ and inhomogeneous stress fields are thus generated within the vicinity of the cavity. Note that for a neo-Hookean medium, we require only $p_{\scriptsize{\mbox{in}}}/\mu\sim 3.5$ to generate a radius that is $20$ times larger than its original size and for soft materials we have $\mu\sim 10^5$ Pa. Furthermore we assume that the fluid inside the cavity is inviscid with zero shear modulus so that the appropriate boundary condition for the incoming antiplane shear wave is traction free.

We show that any object placed inside the inviscid fluid in the inflated cavity is cloaked in the sense that for a neo-Hookean medium,  scattering coefficients associated with the inflated cavity (with $Ka\sim O(1)$) are \textit{exactly} the same as those associated with the undeformed medium (i.e.\ scattering is weak). In this sense the proposed scheme has connections with reduction of scattering by coating regions with special materials as has been discussed in Alu and Engheta (2005) although as we stress again, no special materials are required here. The effect in the pre-stressed configuration is that waves are bent around the large cylindrical cavity by the induced anisotropy and inhomogeneity and so an object placed inside this region is not seen by an observer located an appreciable distance away. Here we also mention the recent work by Amirkhizi et al.\ (2010) who also discuss the ability to guide waves by continuous changes in anisotropy induced by a continuous change of microstructural properties.

We have non-perfect cloaking in the sense that the original cavity is of finite size. This is analogous to the fact that in the classical (linear) cloaking coordinate transformation, perfect cloaking is not possible because of the singular mapping. However here we note that we may always choose this cavity size to ensure that scattering is as weak as possible. Furthermore and perhaps most importantly, the inhomogeneous moduli and cloaking material are generated naturally by the use of nonlinear pre-stress. The continuous change in inhomogeneity induced by pre-stress also means that dispersive effects are not present, as they would be by the use of inhomogeneous metamaterials for example. This property is beneficial for broadband cloaking.

We note that nonlinear elastic pre-stress has been increasingly considered in recent times in order to control elastic waves, notably in order to tune stop and pass bands of composite materials (Parnell 2007, Bertoldi and Boyce 2008, Bigoni et al.\ 2008, Gei et al.\ 2009).

In what follows, in section \ref{sec:classical} we consider the classical linear coordinate transformation used in order to achieve cloaking in the antiplane elastic wave context. In section \ref{sec:prestress} we introduce the theory of nonlinear elasticity required in order to solve the pre-stress problem, and thus determine an equation which links the internal pressure inside the cavity to its deformed radius. Section \ref{sec:incremental} then describes the incremental wave equation which results from considering linearizations about the pre-stressed state. This resulting incremental equation is the equation which has equivalent form to the standard equation in cloaking metamaterials. In section \ref{sec:explicit} we show how an explicit solution to this incremental equation may be obtained thus illustrating that cloaking is theoretically possible by the use of pre-stress. We conclude in section \ref{sec:conc} indicating possibilities for the general elastodynamic case. Some details are located in appendices for ease of reading.

\section{Classical coordinate transformation for antiplane waves} \label{sec:classical}

Take an unbounded homogeneous elastic material with shear modulus $\mu$ and density $\rho$ and introduce a Cartesian coordinate system $(X,Y,Z)$.
%Consider elastic antiplane wave propagation in this medium with the wave polarized in the $Z$ direction and propagating in the $XY$ plane. This is a two-dimensional problem because it is invariant under translations along the $Z$ axis.
Let us also consider a cylindrical polar coordinate system $(R,\Theta,Z)$ with some origin $\mathbf{O}$, whose planar variables are related to the Cartesian coordinate system in the usual form $X=R\cos\Theta, Y=R\sin\Theta$. Suppose that there is a time-harmonic line source, polarized in the $Z$ direction and located at $(R_0,\Theta_0)$, with circular frequency $\omega$ and amplitude $C$ (which is a force per unit length in the $Z$ direction). This generates time-harmonic elastic antiplane waves with displacement  $\mathbf{U}=\Re[W(X,Y)\mathbf{e}_Z\exp(-i\om t)]$ where $\mathbf{e}_Z$ is a unit basis vector in the $Z$ direction. The displacement $W$ is governed by
\begin{align}
\nabla_{\mathbf{X}}\cdot\left(\mu\nabla_{\mathbf{X}} W\right) + \rho\om^2 W &= \frac{C}{R_0}\de(R-R_0)\de(\Theta-\Theta_0) \label{antiplane}
\end{align}
where $\nabla_{\mathbf{X}}$ indicates the gradient operation in the ``untransformed'' frame.

The coordinate transformation theory of cloaking concerns itself with finding transformations $\chi$ to new coordinates $\mathbf{x}=\chi(\mathbf{X})$ which preserve the governing equation of motion but result in modifications to the material properties $\mu$ and $\rho$. This then allows waves to be guided around specific regions of space. The transformation can be quite general, the importance being that it possesses some singularity (Norris 2008). The classical linear mapping for a cloak for antiplane waves (cf.\ acoustics) expressed in plane cylindrical polar coordinates, takes the form (Pendry et al.\ 2006)
\begin{align}
r &= R_1 +R\left(\frac{R_2-R_1}{R_2}\right), & \theta &= \Theta, & z&=Z, \label{classicaltrans}
\end{align}
for $0\leq R\leq R_2$ and the identity mapping for all $R>R_2$ for some chosen $R_1,R_2\in\mathbb{R}$ such that $R_2<R_0$ i.e.\ the line source remains outside the cloaking region. The cloaking region is thus defined by $r\in[R_1,R_2]$.

%where $R$ and $r$ are the untransformed and transformed radial coordinate respectively. This has the problem that it is singular at $R=0$.
%For a homogeneous material with shear modulus $\mu$ and density $\rho$, time-harmonic anti-plane waves with displacement field $\mathbf{u}=(0,0,w(x,y))$ are governed by the standard wave equation
%\begin{align}
%\mu\nabla^2 w + \rho\om^2 w &= \frac{C}{\mu r_0}\de(r-R_0)\de(\theta-\Theta_0),
%\end{align}
%where the line source is thus of strength (force per unit length) $C$.

As is evident from \eqref{classicaltrans} we use upper and lower case variables for the untransformed and transformed problems respectively. Therefore under this mapping the form of the governing equation \eqref{antiplane} remains unchanged for $R=r>R_2$, i.e.\ we have
\begin{align}
\nabla_{\mathbf{x}}\cdot\left(\mu\nabla_{\mathbf{x}} w\right) + \rho\om^2 w &= \frac{C}{R_0}\de(r-R_0)\de(\theta-\Theta_0) \label{antiplane2}
\end{align}
whereas for $0\leq R\leq R_2$, corresponding to the transformed domain $R_1\leq r\leq R_2$, the transformed equation takes the form (in transformed cylindrical polar coordinates $r,\theta=\Theta$)
\begin{align}
\frac{1}{r}\deriv{}{r}\left(r\mu_r(r)\deriv{w}{r}\right) + \frac{\mu_{\theta}(r)}{r^2}\derivtwo{w}{\theta} + d(r)\om^2 w &= 0 \label{inhomwave}
\end{align}
where
\begin{align}
\mu_r(r) &= \mu\frac{(r-R_1)}{r}, & \mu_{\theta}(r) &= \mu\frac{r}{(r-R_1)}, & d(r) &= \rho\frac{(r-R_1)}{r}\left(\frac{R_2}{R_2-R_1}\right)^2. \label{classicalprops}
\end{align}
Equation \eqref{inhomwave} is homogeneous since the line source is located in the outer domain. Note rather importantly that two effects are thus necessary for cloaking of this form. Firstly \textit{both} the shear modulus and density must be inhomogeneous. Secondly the shear modulus must be anisotropic. Finally we note that material properties of this form cannot be constructed exactly since the shear modulus $\mu_{\theta}$ becomes unbounded as $r\rightarrow R_1$ (the inner boundary of the cloak) whilst the density tends to zero in this limit.

In figure \ref{fig2} we show an example of antiplane cloaking with a cloaking region defined by inner and outer radii $R_1=2\pi, R_2=4\pi$ and the line source located at $R_0=8\pi, \Theta_0=0$. This figure is plotted by solving \eqref{antiplane2}-\eqref{classicalprops} which is achieved by noting that the solution to \eqref{antiplane} is (see \ref{unstressed})
\begin{align}
W(R,\Theta) &= W_i(R,\Theta) = \frac{C}{4i\mu}\tn{H}_0(KS)
\end{align}
where $S=\sqrt{(X-X_0)^2+(Y-Y_0)^2}$ and where $\tn{H}_0(KR)=\tn{H}_0^{(1)}(KR)=\tn{J}_0(KR)+i\tn{Y}_0(KR)$ is the Hankel function of the first kind and order zero, noting that $\tn{J}_0$ and $\tn{Y}_0$ are Bessel functions of the first and second kind respectively, also of order zero. We note that this solution retains the same form in the mapped configuration for $R>R_2$ since the mapping in this region is the identity mapping. Therefore for $r>R_2$ we have
\begin{align}
w(r,\theta) &= \frac{C}{4i\mu}\tn{H}_0(Ks)
\end{align}
where $s=\sqrt{(x-X_0)^2+(y-Y_0)^2}$, noting that the source location remains at $x=X_0=R_0\cos\Theta_0, y=Y_0=R_0\sin\Theta_0$. For $0\leq R\leq R_2$, corresponding to $R_1\leq r \leq R_2$, we use Graf's addition theorem with $R_0>R_2$ (see \eqref{H0scat}) followed by the mapping \eqref{classicaltrans} so that in the region $R_1\leq r\leq R_2$
\begin{align}
w(r,\theta) &= \frac{C}{4i\mu}\sum_{n=-\infty}^{\infty} \tn{H}_n(KR_0)\tn{J}_n(KR)e^{in(\Theta-\Theta_0)}
\end{align}
where here $R=R_2(r-R_1)/(R_2-R_1)$ and $\Theta=\theta$.

\begin{figure}[h!]
\begin{center}
\includegraphics[scale=0.6]{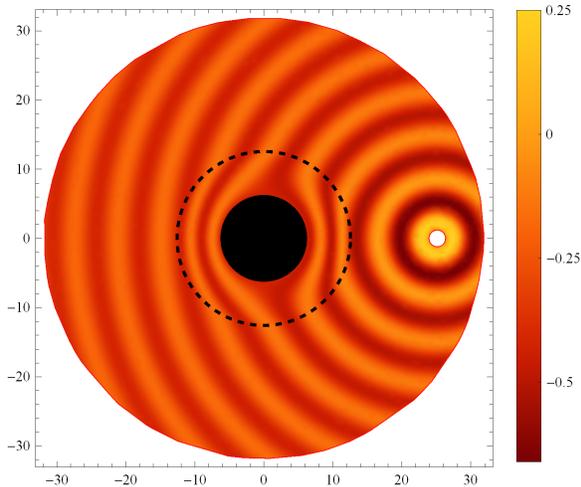}
\caption{A region of (nondimensionalized) radius $Ka=2\pi$ is cloaked by the use of an elastic cloak in $2\pi\leq Kr\leq 4\pi$ whose properties \eqref{classicalprops} are obtained by applying the classical linear transformation \eqref{classicaltrans} to Helmholtz equation in two dimensions. Here the line source is located at $Kr=KR_0=8\pi$ and $\Theta_0=0$ and is shown as a small finite region to avoid the large values near its origin.}
\label{fig2}
\end{center}
\end{figure}

As discussed in the introduction, cloaking thus requires very special materials having properties such as \eqref{classicalprops} above and the construction of such materials is non-trivial. In what follows we shall show that a similar cloaking effect can be obtained as that described above but with naturally obtained anisotropic, inhomogeneous shear moduli and a \textit{scalar, homogeneous} density, thus not requiring the construction of special materials. This effect is due to nonlinear elastic pre-stress. The corresponding equation still has the form of \eqref{inhomwave} but now with
\begin{align}
\mu_r(r) &= \mu\left(\frac{r^2+M}{r^2}\right), & \mu_{\theta}(r) &= \mu\left(\frac{r^2}{r^2+M}\right), & d &= \rho,
\end{align}
where $M=A^2-a^2$ relates the undeformed and deformed radii of the cylindrical cavities $A$ and $a$ respectively before and after pre-stress. We note further that the density remains uniform.

\section{Pre-stress} \label{sec:prestress}

Let us consider an isotropic \textit{incompressible} neo-Hookean material of infinite extent with a cylindrical cavity of initial radius $A$ at the origin. The constitutive behaviour of a neo-Hookean material is described by the strain energy function (Ogden 1997)
\begin{align}
\mathcal{W} &= \frac{\mu}{2}(I_1-3) = \frac{\mu}{2}(\la_1^2+\la_2^3+\la_3^3 - 3) \label{NHSEF}
\end{align}
where $I_j$ and $\la_j, j=1,2,3$ are the principal strain invariants and principal stretches respectively, of the deformation which is to ensue. Stresses associated with the deformation are obtained via derivatives of this strain energy function, and are defined in \eqref{Cauchy stress tensor} below. We consider the initial deformation as depicted on the left of figure \ref{fig2} which arises due to an increase in internal pressure inside the cavity region, leading to a cavity with larger radius. In order to describe this deformation mathematically we therefore write
\begin{equation}
R=R(r),\qquad \Theta=\theta,\qquad Z=z,
\label{initial deformation}
\end{equation}
where $(R,\Theta,Z)$ and $(r,\theta,z)$ are cylindrical polar coordinates in the undeformed and deformed configurations, denoted B and b respectively. The function $R(r)$ will be determined from the radial equation of equilibrium and incompressibility condition below. Note the convention introduced in \eqref{initial deformation}, i.e.\ that upper case variables correspond to the undeformed configuration whilst lower case corresponds to the deformed configuration. This notation is therefore analogous to that used for the untransformed and transformed coordinates for classical cloaking as described in section \ref{sec:classical}. Since we are interested in incremental perturbations from the pre-stressed state, it will be convenient for us to derive equations in terms of coordinates in the deformed configuration. Position vectors in the undeformed (upper case) and deformed (lower case) configurations are respectively
\begin{align}
\mathbf{X}&=\left(
\begin{array}{c}
R\cos\Theta \\
R\sin\Theta \\
Z
\end{array}
\right)=\left(
\begin{array}{c}
R(r)\cos\theta \\
R(r)\sin\theta \\
z
\end{array}
\right), &
\mathbf{x}&=\left(
\begin{array}{c}
r\cos\theta \\
r\sin\theta \\
z
\end{array}
\right).
\end{align}

\begin{figure}[h!]
\begin{center}
\psfrag{X}{$X$}
\psfrag{Y}{$Y$}
\psfrag{R0}{$R_0$}
\psfrag{T0}{$\Theta_0$}
\psfrag{p}{$p_{\scriptsize{\mbox{in}}}=0$}
\psfrag{p1}{$p_{\scriptsize{\mbox{in}}}>0$}
\includegraphics[scale=0.4]{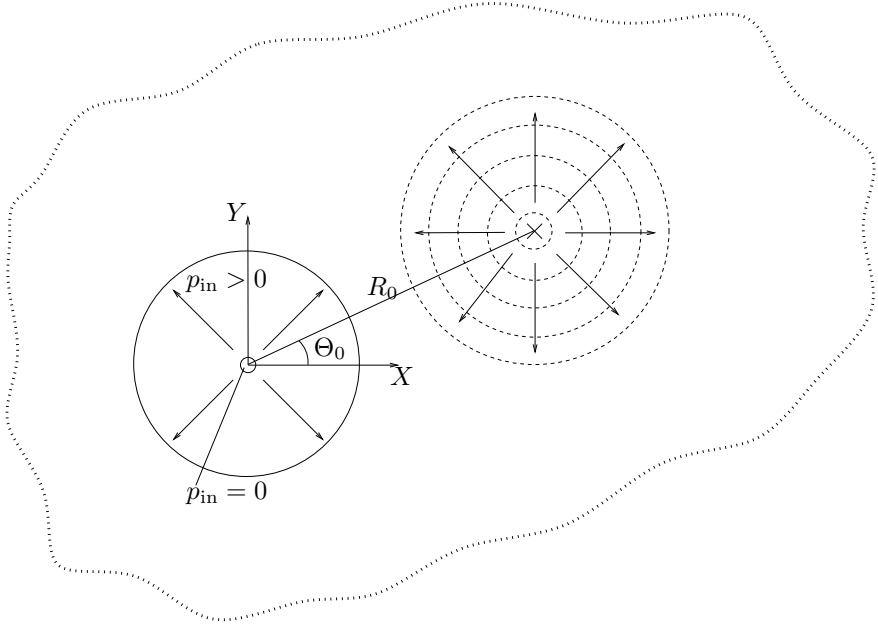}
\caption{In an unbounded medium, a time-harmonic line source, located at $(R_0,\Theta_0)$ is indicated by crossed lines and we show schematically the direction of outward propagating waves and wave-fronts. This wave is scattered from a small cavity of radius $A\ll\Lambda$ where $\Lambda$ is the wavelength of the source, with internal pressure $p_{\scriptsize{\mbox{in}}}=0$ in the undeformed configuration. When the pressure is increased, $p_{\scriptsize{\mbox{in}}}>0$, the cavity increases its radius to $a$ such that $a/\Lambda=O(1)$. A region of inhomogeneous deformation develops around the deformed cavity.}
\label{fig2}
\end{center}
\end{figure}

Using (\ref{initial deformation}), it can be shown that the principal stretches for this deformation in the radial, azimuthal and longitudinal directions, respectively, are
\begin{align}
\la_1=\lambda_r&=\frac{dr}{dR}=\frac{1}{R^\prime(r)}, & \la_2&=\lambda_\theta=\frac{r}{R(r)}, & \lambda_z&=1.
\end{align}
The deformation gradient tensor $\mathbf{F}$ is defined by
\begin{equation}
\mathbf{F}=\Grad\mathbf{x},
\label{FGradx}
\end{equation}
where $\Grad$ represents the gradient operator in $\mathbf{X}$ and here we have
\begin{equation}
\mathbf{F}=\left(\begin{array}{ccc}
\lambda_r & 0 & 0 \\
0 & \lambda_\theta & 0 \\
0 & 0 & \lambda_z
\end{array}\right)=\left(\begin{array}{ccc}
(R^\prime(r))^{-1} & 0 & 0 \\
0 & r/R(r) & 0 \\
0 & 0 & 1
\end{array}\right).
\end{equation}
For an incompressible material, we must have $J=\det\mathbf{F}=1$, and so
\begin{equation}
\lambda_r\lambda_\theta\lambda_z=\frac{r}{R(r)R^\prime(r)}=1. \label{ode}
\end{equation}
The differential equation \eqref{ode} can easily be solved to yield
\begin{equation}
R(r)=\sqrt{r^2+M}, \label{eqn:Q(r)}
\end{equation}
where $M$ is a constant, the form of which is defined by the corresponding radii of cavity boundaries, i.e.
\begin{equation}
M=A^2-a^2, \label{alpha}
\end{equation}
with the equation linking $a, A$ and internal pressure $p_{\tn{in}}$ given shortly.

From Ogden (1997, 2007), the Cauchy and nominal stress tensors for an incompressible material are respectively given by
\begin{align}
\mathbf{T} &= \mathbf{F}\frac{\partial \mathcal{W}}{\partial\mathbf{F}}+Q\mathbf{I}, &  \mathbf{S} &= \frac{\partial \mathcal{W}}{\partial\mathbf{F}}+Q\mathbf{F}^{-1}
\label{Cauchy stress tensor}
\end{align}
where $\mathcal{W}$ is the neo-Hookean strain energy function introduced in \eqref{NHSEF}, $\mathbf{I}$ is the identity tensor and $Q$ is the scalar Lagrange multiplier associated with the incompressibility constraint. Note that the derivative with respect to $\mathbf{F}$ is defined component-wise by
\begin{align}
\left\{\deriv{\mathcal{W}}{\mathbf{F}}\right\}_{ij} = \deriv{\mathcal{W}}{F_{ji}}
\end{align}
which is the convention introduced in Ogden (1997).
%Equation (\ref{Cauchy stress tensor}) can be written in components by
%\begin{equation}
%T_{ij}=F_{i\alpha}\frac{\partial \mathcal{W}}{\partial F_{j\alpha}}+Q\delta_{ij},
%\end{equation}
%where $\delta_{ij}$ is the Kronecker delta.
The tensors $\mathbf{F}$ and $\mathbf{T}$ are diagonal, the non-zero components of the latter being given by
\begin{align}
T_{rr}&= \mu\left(\frac{r^2+M}{r^2}\right)+Q,  &
T_{\theta\theta}&= \mu\left(\frac{r^2}{r^2+M}\right)+Q, &
T_{zz}&=\mu+Q. \label{cauchycomps}
\end{align}
The static equations of equilibrium are
\begin{equation}
\Div\mathbf{T}=\mathbf{0},
\label{static equations of equilibrium}
\end{equation}
where $\Div$ signifies the divergence operator in $\mathbf{x}$. These reduce to
\begin{align}
\frac{\partial T_{rr}}{\partial r}+\frac{1}{r}(T_{rr}-T_{\theta\theta})&=0, &
\frac{\partial T_{\theta\theta}}{\partial\theta}&=0, &
\frac{\partial T_{zz}}{\partial z}&=0. \label{eoms}
\end{align}
The second and third of these simply yield $Q=Q(r)$. Integrating the first of \eqref{eoms}, using $T_{rr}$ and $T_{\theta\theta}$ from \eqref{cauchycomps} and applying the traction condition $\mathbf{t}|_{r=a}=\mathbf{T}\mathbf{n}|_{r=a}=p_{\scriptsize{\mbox{in}}}\mathbf{e}_r$ where $\mathbf{e}_r$ is the unit vector in the radial direction and $\mathbf{n}=-\mathbf{e}_r$ is the (outward) unit normal from the elastic material, so that $T_{rr}|_{r=a}=-p_{\scriptsize{\mbox{in}}}$, we find that the radial stress is determined by
\begin{align}
T_{rr}(r)+p_{in} &= \frac{\mu}{2}\left(M\left(\frac{1}{r^2}-\frac{1}{a^2}\right)+\log\left(\frac{r^2+M}{a^2+M}\right)-\log\left(\frac{r^2}{a^2}\right)\right). \label{Trreqn}
\end{align}
This allows $Q$ to be determined from \eqref{cauchycomps}.
%\begin{align}
%\frac{\partial Q}{\partial r} &= \frac{\mu}{L}\frac{M^2}{r^3(r^2+M)}, &
%\frac{\partial Q}{\partial\theta}&=\frac{\partial Q}{\partial z}=0.
%\label{eqn:governingequations}
%\end{align}
%Hence $Q$ depends only on $r$ and is determined, up to an arbitrary constant by
%\begin{align}
%Q %& = \int\frac{M^2}{Lr^3(M+r^2)}\mu dr\nonumber\\
%& = \frac{\mu}{2L}\left(\log\left(\frac{r^2+M)}{r^2}\right)-\frac{M}{r^2}\right).
%\label{eqn:p}
%\end{align}

\begin{figure}[h!]
\begin{center}
\psfrag{p}{$p_{\scriptsize{\mbox{in}}}/\mu$}
\psfrag{a}{$a/A$}
\includegraphics[scale=0.6]{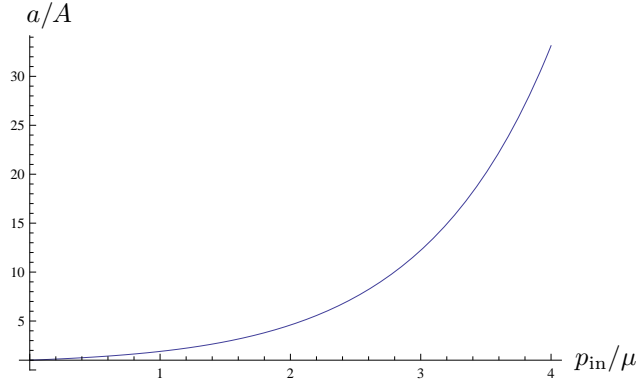}
\caption{Plot of the deformed to undeformed radius ratio $a/A$ as a function of $p_{\scriptsize{\mbox{in}}}/\mu$.} \label{radius}
\end{center}
\end{figure}

Next using the condition that $T_{rr}\rightarrow 0$ as $r\rightarrow\infty$ in \eqref{Trreqn} yields the relationship
\begin{align}
\frac{p_{\scriptsize{\mbox{in}}}}{\mu} &= \frac{1}{2}\left(1-\frac{A^2}{a^2}+\log\left(\frac{a^2}{A^2}\right)\right). \label{pcondition}
\end{align}
This nonlinear equation determines the deformed to undeformed radius ratio $a/A$ as a function of the scaled pressure $p_{\scriptsize{\mbox{in}}}/\mu$ and we plot this relationship in figure \ref{radius}, noting that moderate values of $p_{\scriptsize{\mbox{in}}}/\mu$ can lead to relatively large values of $a/A$. We note that for rubbery materials, the shear modulus is usually low, $\mu=O(10^5)$ Pa.

\section{Incremental deformations} \label{sec:incremental}

The antiplane line source will generate outgoing waves and subsequent scattered waves from the cavity. We now consider how to obtain the (incremental) equation which governs wave propagation in the pre-stressed medium. This equation is derived by appealing to the theory of ``small-on-large'' (Ogden 1997, 2007). To this end let us consider incremental deformations from the (statically) deformed body $b$. We consider a finite deformation of the original body $B$ to a new deformed state $\bar{b}$ which is close to the configuration $b$. The position vector in the new deformed state $\bar{b}$ is defined by $\bar{\mathbf{x}}$ and we define
\begin{align}
\mathbf{u} &= \bar{\mathbf{x}}-\mathbf{x}, & |\mathbf{u}|\ll 1
\end{align}
as the difference between position vectors in $\bar{b}$ and $b$. The term $\mathbf{u}$ is called an \textit{incremental displacement} since $\bar{b}$ is close to $b$. Since our line source is of antiplane type and is time-harmonic, we therefore assume that the total incremental deformation is of the form
\begin{align}
\mathbf{u} &= \Re[w(x,y)\mathbf{e}_z\exp(-i\om t)],
\end{align}
where $\mathbf{e}_z$ is a unit basis vector in the $z$ direction. On employing the theory of small-on-large, the technical details of which can be found in \ref{app}, we find that the governing equation of motion for $w$ is
%is
%\begin{align}
%\frac{1}{r}\deriv{}{r}\left(r\zeta_{rz}\right)+\frac{1}{r}\deriv{\zeta_{\theta z}}{\theta}+\rho\om^2 w = \frac{c}{r_0}\de(r-r_0)\de(\theta-\theta_0) \label{eomzeta}
%\end{align}
%where $\zeta_{ij}$ are components of the (non-symmetric) stress tensor commonly referred to as the \textit{push-forward of the incremental nominal stress}. For reference it is related to the (perhaps more physically intuitive and symmetric) incremental Cauchy stress by the relation $\boldsymbol{\tau} = \boldsymbol{\zeta} + \boldsymbol{\ga}\mathbf{T}$ where $\boldsymbol{\ga}=\grad\mathbf{u}$ and $\grad$ refers to the gradient with respect to $\mathbf{x}$.
%
%In the case under consideration here, the appropriate components $\zeta_{rz}=\tau_{rz}$ and $\zeta_{r\theta}=\tau_{r\theta}$ (this is not the case generally) and these stresses are defined in terms of the antiplane displacement $w$ by
%\begin{align}
%\tau_{rz} = \zeta_{rz} &= \mu_r(r)\frac{\pa w}{\pa r}, & \tau_{\theta z} = \zeta_{\theta z} &= \mu_{\theta}(r)\frac{1}{r}\frac{\pa w}{\pa \theta}. \label{eomzetarel}
%\end{align}
%Here we have defined the anisotropic shear moduli
%\begin{align}
%\mu_r(r) &= \frac{\mu}{L}\left(\frac{r^2+M}{r^2}\right), & \mu_{\theta}(r) &= \frac{\mu}{L}\left(\frac{r^2}{r^2+M}\right). \label{murmut}
%\end{align}
%Using \eqref{eomzetarel} in \eqref{eomzeta} gives the governing equation of motion in terms of $w$:
\begin{align}
\frac{1}{r}\deriv{}{r}\left(r\mu_r(r)\deriv{w}{r}\right)+\frac{\mu_{\theta}(r)}{r^2}\derivtwo{w}{\theta}+\rho\om^2 w &= \frac{C}{r_0}\de(r-r_0)\de(\theta-\theta_0) \label{incr}
\end{align}
where
\begin{align}
\mu_r(r) &= \mu\left(\frac{r^2+M}{r^2}\right), & \mu_{\theta}(r) &= \mu\left(\frac{r^2}{r^2+M}\right)
\end{align}
and $C$ is the force per unit length of the line source (as associated with the original line source). We note that as $r\rightarrow\infty$, \eqref{incr} reduces to
\begin{align}
\frac{1}{r}\deriv{}{r}\left(r\deriv{w}{r}\right)+\frac{1}{r^2}\derivtwo{w}{\theta}+K^2 w &= \frac{C}{\mu r_0}\de(r-r_0)\de(\theta-\theta_0)
\end{align}
where $K^2=\rho\om^2/\mu$ is the wavenumber associated with the undeformed material. Note that as $r\rightarrow\infty$ the deformation does not affect the wave and the equation is simply the scalar wave equation in the far field.

\section{Cloaking via pre-stress} \label{sec:explicit}

\subsection{Explicit solution of the incremental equation}

In \ref{unstressed} we derive the standard form of the scattered antiplane elastic wave field \eqref{Wsfld}-\eqref{An1} which arises due to scattering from a cylindrical cavity (filled with inviscid fluid) in an \textit{unstressed} medium, with a time-harmonic line source located at $(R_0,\Theta_0)$. In the undeformed configuration $B$ we take $KA\ll 1$ and therefore the leading order scattering coefficient is $O((KA)^2)$ (Parnell and Abrahams 2010), i.e.\ the scattered field is weak compared with the $O(1)$ incident field. When $KA=O(1)$ this scattered field is no longer weak in general.

Let us now consider the pre-stressed problem in the deformed configuration $b$. The line source is thus moved to the location $(r_0,\theta_0)=(\sqrt{R_0^2-M},\Theta_0)$ (this movement will itself be negligible provided it is far enough away from the cavity) and we retain the strength $C$ of the line source associated with the undeformed configuration. Let us consider pressures $p_{\scriptsize{\mbox{in}}}$ that are large enough in order to inflate the cavity to a radius $a$ such that $Ka=O(1)$.

%
%We now consider the analogous problem to \eqref{inhomprob0} in the \textit{deformed} configuration. We take a line source on the right hand side of \eqref{incr} with magnitude $c=C$, i.e.\
%\begin{align}
%\frac{\mu}{L}\left[\left(1+\frac{M}{r^2}\right)\derivtwo{w}{r}+\frac{1}{r}\left(1-\frac{M}{r^2}\right)\deriv{w}{r}+\frac{1}{(r^2+M)}\derivtwo{w}{\theta}\right]+\rho\om^2 %w &= \frac{C}{L r_0}\de(r-r_0)\de(\theta-\theta_0). \label{incMRw2}
%\end{align}

The equation governing the incremental waves in the configuration $b$ is \eqref{incr}, or explicitly
\begin{align}
\frac{1}{r}\deriv{}{r}\left(\left(r+\frac{M}{r}\right)\deriv{w}{r}\right)+\frac{1}{(r^2+M)}\derivtwo{w}{\theta}+K^2 w &= \frac{C}{\mu r_0}\de(r-r_0)\de(\theta-\theta_0), \label{incMRw2}
\end{align}
noting again that $K^2=\rho\om^2/\mu$ is the wavenumber associated with the undeformed material. In this specific case of a neo-Hookean elastic material we can, in fact, solve this problem analytically and we show, perhaps surprisingly, that the scattering coefficients are completely unaffected by the pre-stress and therefore even though the cavity has increased significantly in size, its scattering effect on incident waves remains weak.

In \eqref{incMRw2}, we introduce the mapping
\begin{align}
R^2 &= r^2+M, & \Theta=\theta \label{mapping}
\end{align}
and we note that this mapping corresponds exactly to the initial finite deformation \eqref{initial deformation} and \eqref{eqn:Q(r)}. Upon defining $W(R,\Theta)=w(r(R),\theta(\Theta))$, we find
\begin{align}
\derivtwo{W}{R}+\frac{1}{R}\deriv{W}{R}+\frac{1}{R^2}\derivtwo{W}{\Theta}+K^2 W = \frac{C}{\mu}\frac{1}{r_0}\de(r-r_0)\de(\Theta-\Theta_0). \label{incMRw4}
\end{align}
%In order to write the right-hand-side of \eqref{incMRw4} in terms of $R$ and $R_0$, note that from \eqref{mapping}
%\begin{align}
%\frac{1}{r_0} &= \int_{0}^{\infty}{\frac{1}{r_0}\de(r-r_0)}{\hspace{0.1cm}dr} = \frac{R_0}{Lr_0^2}\int_{0}^{\infty}{\de(r-r_0)}{\hspace{0.1cm}dR}
%\end{align}
%and therefore
%\begin{align}
%\int_{0}^{\infty}{\frac{1}{Lr_0}\de(r-r_0)}{\hspace{0.1cm}dR} = \frac{1}{R_0} = %\int_{0}^{\infty}{\frac{1}{R_0}\de(R-R_0)}{\hspace{0.1cm}dR}
%\end{align}
%so that
It is straightforward to show that
\begin{align}
\frac{1}{r_0}\de(r-r_0) &= \frac{1}{R_0}\de(R-R_0)
\end{align}
and hence \eqref{incMRw4} becomes
\begin{align}
\nabla^2 W + K^2 W = \frac{C}{\mu}\frac{1}{R_0}\de(R-R_0)\de(\Theta-\Theta_0). \label{incMRw5}
\end{align}
We conclude that the solution of \eqref{incMRw5} is entirely equivalent to the solution of \eqref{inhomprob} obtained in \ref{unstressed}. The incident field is therefore given by \eqref{w0scat} and \eqref{H0scat}, which thus gives rise to the scattered field associated with \eqref{incMRw5} as (cf.\ \eqref{Wsfld})
\begin{align}
W_s(R) &= \sum_{n=-\infty}^{\infty}(-i)^n a_n\tn{H}_n(KR)e^{in(\Theta-\Theta_0)}. \label{WsR}
\end{align}
Referring to \ref{unstressed}, and imposing the necessary traction free boundary condition on $R=A$, the scattering coefficients $a_n$ are given by
\begin{align}
a_n &= A_n = \frac{C(-1)^n}{4\mu i^{n-1}}\frac{\tn{J}_n'(KA)}{\tn{H}_n'(KA)}\tn{H}_n(KR_0). \label{an}
\end{align}

We now map back to the deformed configuration, using \eqref{mapping} in \eqref{H0scat} and \eqref{WsR}. The incident and scattered fields in the deformed configuration are therefore
\begin{align}
w_i(r) &=
\frac{C}{4i\mu}\times
\begin{cases}
\sum_{n=-\infty}^{\infty} \tn{H}_n(K\sqrt{r_0^2+M})\tn{J}_n(K\sqrt{r^2+M})e^{in(\theta-\theta_0)}, & r<r_0, \\
\sum_{n=-\infty}^{\infty} \tn{H}_n(K\sqrt{r^2+M})\tn{J}_n(K\sqrt{r_0^2+M})e^{in(\theta-\theta_0)}, & r>r_0,
\end{cases} \label{wincr}
\end{align}
and
\begin{align}
w_s(r) &= \sum_{n=-\infty}^{\infty}(-i)^n a_n\tn{H}_n\left(K\sqrt{r^2+M}\right)e^{in(\theta-\theta_0)} \label{wssol}
\end{align}
where we used Graf's addition theorem to obtain the form \eqref{wincr} (Martin 2006).

Perhaps most important to note is that the scattering coefficients $a_n$ depend on the \textit{initial} distance $R_0$ between the centre of the cavity and the source location, and the \textit{undeformed} cavity radius $A$ and they are unaffected by the pre-stress, $a_n=A_n$. The wave field close to the cavity is affected by the pre-stress and in fact waves are bent around the object, whereas in the far-field the wave that is seen will be identical to the field that would be seen in that region when scattering takes place from the undeformed cavity. Since scattering is weak in the latter case ($KA\ll 1$), it will also be weak from the deformed (large) cavity ($Ka=O(1)$).

The analytical solution to the scattering problem above is used in Parnell and Abrahams (2011) to consider antiplane wave scattering in a pre-stressed microvoided composite with a neo-Hookean host phase.

\subsection{Cloaking}

In order to illustrate the above theory, consider the case where $a/A=20$ with $KA=2\pi/20, Ka=2\pi$. We set $KR_0=8\pi, \Theta_0=0$. As in figure \ref{fig2}, on introducing the wavelength $\Lambda$ via $\Lambda=2\pi/K$, this choice ensures that $\Lambda/A=20$ (weak scattering) whereas $\Lambda/a=1$. Having the ratio $\Lambda/a=1$ in an unstressed medium would normally result in strong scattering and the notable presence of a scattered field and shadow region. Using pre-stress however, the scattering effect of a cylindrical cavity of radius $a$ such that $Ka=O(1)$ is equivalent to that of the undeformed cavity with $KA\ll 1$, i.e.\ very weak. Note that an inner pressure of around $p_{\scriptsize{\mbox{in}}}/\mu=3.5$ ensures this deformation.

\begin{figure}[h!]
\begin{center}
\includegraphics[scale=0.5]{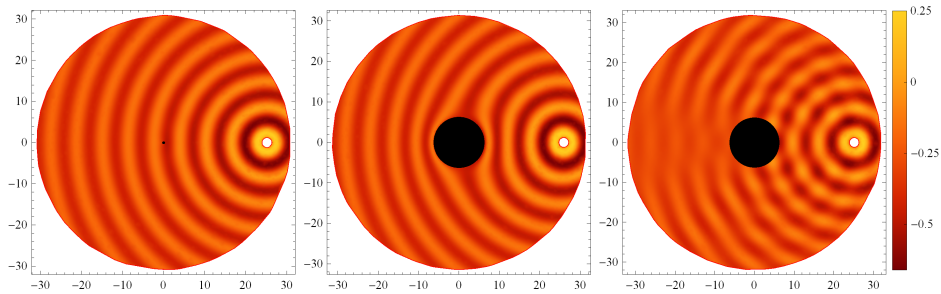}
\caption{Total antiplane wave-field due to a line source located at $R_0=8\pi, \Theta_0=0$ in an undeformed configuration. The source is shown as a small finite region to avoid the large values near its origin. We show the total field in the case of a small cavity with $KA=2\pi/20$ (left), a deformed cavity in the pre-stressed medium ($A\rightarrow a$) with $Ka=2\pi$ (middle) and a large cavity with $KA=2\pi$ in an unstressed medium (right). This illustrates the possibility of cloaking via pre-stress.}
\label{totfld}
\end{center}
\end{figure}

We illustrate this situation with three plots of the total wave field in figure \ref{totfld}. On the left we show the total field $W=W_i+W_s$ (defined in \eqref{H0scat} and \eqref{Wsfld}) resulting from scattering from the small cylindrical cavity ($KA=2\pi/20$) in an undeformed medium. Since $\Lambda\gg A$, scattering is very weak and the scattering effect cannot be seen on the plot. In the middle plot we show the equivalent problem but with the pre-stressed cavity inflated from $A$ to $a$ so that $Ka=2\pi$ and the total displacement field plotted is $w=w_i+w_s$ as defined in \eqref{wincr} and \eqref{wssol}. As shown theoretically above, the scattered field remains the same as for the undeformed problem and therefore even though $Ka=2\pi=O(1)$, scattering is weak. This is in contrast to the plot on the right of the figure where we show the total field wave $W=W_i+W_s$ for an undeformed cavity with $KA=2\pi$, illustrating the total wave field that \textit{would} result in an unstressed medium in this scenario and the fact that it gives rise to strong scattering and a noticable shadow region.

%
%\section{More general materials}
%
%The above mapping works only for neo-Hookean materials and not for more general constitutive behaviour such as neo-Hookean, Gent, Fung, etc. In those cases what should we do? Well consider the following. Suppose that the initial configuration with small cavity now case an inhomogeneous elastic modulus, as required in the classical cloaking case but for now let us treat this as unknown, but we still insist that the density is homogeneous stating the dynamic response in this coordinate system as
%\begin{align}
%\frac{1}{R}\deriv{}{R}\left(R\hat{\mu}_R(R)\deriv{W}{R}\right)+\frac{\hat{\mu}_{\Theta}(R)}{R^2}\derivtwo{W}{\Theta}+\rho\om^2 W&= 0.
%\end{align}
%Now, knowing that the incremental equation
%
%
%

\section{Conclusions} \label{sec:conc}

In this article we have shown how cloaking of objects from antiplane elastic waves can be achieved by nonlinear elastic pre-stress of a neo-Hookean material. In this case the deformation leaves the scattering coefficients unchanged and thus provided the initial cavity is a weak scatterer, the deformed (large) cavity will also be a weak scatterer. In particular we note that the inhomogeneous anisotropic shear modulus that is found in the incremental equations arises naturally by virtue of the pre-stress. This means that no special metamaterials are required. Under this deformation the density also remains isotropic and homogeneous. That this was possible was due to the special transformation property of the incremental equations for neo-Hookean materials, in that the mapping which takes them back to the Helmholtz equation is exactly the same mapping (physical deformation) associated with the pre-stress.

This same property does not hold for other nonlinear elastic materials (Mooney-Rivlin, Gent, Fung, etc.). This does not necessarily mean that a similar effect could not be achieved with these other materials however. If it can be shown that the scattering coefficients in the deformed configuration are changed only slightly from the scattering coefficients associated with the undeformed (weak scattering) cavity then similar effects would result. One should also consider finite regions to be pre-stressed (to form finite cloaks) which would require compressibility.

Additionally, it appears that the notion of pre-stress permits active control: the size of the obstacle which can be cloaked is limited by the size of the pre-stressed (inflated) cavity. This is modified by changing the magnitude of the pressure inside the cavity. The continuous change in inhomogeneity induced by pre-stress also means that dispersive effects are not present, as they would be by the use of inhomogeneous metamaterials for example. This property is beneficial for broadband cloaking.

We should note that there can be various problems associated with the stability of inflating cavities in elastic media. This issue may therefore require some consideration as to the construction of an appropriate elastic cloaking medium and in particular it may necessitate the use of elastomeric composites which can permit the inflation with greater ease.

Finally, the obvious question is can this idea be generalized to other elastodynamic problems? Cloaking in higher dimensional elastodynamic problems has proved difficult since Navier's equations are not invariant under coordinate transformations (Milton et al.\ 2006). Note that a two dimensional transformation for in-plane elastic waves was considered in Brun et al.\ (2009) which led to invariance of the governing equations except that the transformed elastic modulus tensor did not possess the minor symmetries. Construction of materials with this property is non-trivial and motivated Norris and Shuvalov (2011) to consider the use of Cosserat materials which also do not possess these minor symmetries. Here, however we note that we have not used a coordinate transformation in order to motivate a cloak with inhomogeneous material properties - the mapping which generates such inhomogeneous properies is a \textit{physical deformation} associated with pre-stress. This naturally generates a transformed elastic modulus tensor (the instantaneous modulus tensor $\mathbf{M}$ as defined in \eqref{Minc}) which does not possess the minor symmetries. Therefore, provided we can ensure that the scattering coefficients in the pre-stressed configuration remain weak, even when $Ka=O(1)$ as was the case above, we can achieve approximate cloaking. This clearly motivates studies on more general elastic waves in two and three dimensions. One could also hypothesize that pre-stress of nonlinearly elastic metamaterials could be useful in the cloaking context.

\appendix
\section{Summary of small-on-large theory} \label{app}

For the application of small-on-large theory we follow the classical approach as in Ogden (1997, 2007) with slight modifications to notation. We consider incremental deformations from the deformed body $b$. To this end consider a finite deformation of the original body $B$ to a new deformed state $\bar{b}$ which is close to the configuration $b$. Position vectors in the deformed states $b$ and $\bar{b}$ are defined by $\mathbf{x}$ and $\bar{\mathbf{x}}$ and we define
\begin{align}
\mathbf{u} &= \bar{\mathbf{x}}-\mathbf{x}, & |\mathbf{u}|\ll 1
\end{align}
as the difference between position vectors in $\bar{b}$ and $b$. Since $\bar{b}$ is close to $b$, $\mathbf{u}$ is called an \textit{incremental displacement} which we assume is time-harmonic and of the \textit{antiplane} type, i.e.\
\begin{align}
\mathbf{u} &=\Re[w(x,y)\mathbf{e}_z\exp(-i\om t)] \label{uwSH}
\end{align}
where $\mathbf{e}_z$ is a unit vector in the $z$-direction in the body $b$, generated by a line source of amplitude $C$ in the deformed configuration.

%%%%%%%%%%%%%%%%%%%%%%%%%%%%

Given the deformation gradient \eqref{FGradx}, we can define the additional deformation gradients $\mathbf{f}=\grad \bar{\mathbf{x}}$ and $\overline{\mathbf{F}}=\Grad\bar{\mathbf{x}}=\mathbf{f}\mathbf{F}$ where $\grad$ and $\Grad$ denote the operation of taking the gradient in $\mathbf{x}$ and $\mathbf{X}$ respectively. Therefore
\begin{align}
\boldsymbol{\Gamma} &= \Grad\mathbf{u} = \Grad(\bar{\mathbf{x}}-\mathbf{x}) = (\mathbf{f}-\mathbf{I})\mathbf{F}
\end{align}
and similarly $\boldsymbol{\gamma} = \mathbf{f}-\mathbf{I}$, so that $\mathbf{f}=\mathbf{I}+\boldsymbol{\ga}$ and $\boldsymbol{\Gamma}=\boldsymbol{\ga}\mathbf{F}$.

Furthermore, given the stresses \eqref{Cauchy stress tensor} in the configuration $b$, we can define the corresponding stresses $\bar{\mathbf{T}}=\mathbf{T}+\boldsymbol{\tau}$ and $\bar{\mathbf{S}}=\mathbf{S}+\mathbf{s}$ in $\bar{b}$ where $\boldsymbol{\tau}$ and $\mathbf{s}$ are the incremental Cauchy and nominal stresses respectively. Upon taking a Taylor expansion about the deformation state $\mathbf{F}$, one can show that for an incompressible material
\begin{align}
\mathbf{s} &= \mathbf{L}:\boldsymbol{\gamma}\mathbf{F} +q \mathbf{F}^{-1} - Q \mathbf{F}^{-1}\boldsymbol{\gamma} \label{sinc}
\end{align}
where $q$ is the increment of $Q$ and $\mathbf{L}=\pa^2 \mathcal{W}/\pa\mathbf{F}^2$ is defined componentwise via $L_{ijkl}=\pa^2 \mathcal{W}/\pa F_{ji}\pa F_{lk}$. The colon notation indicates $\mathbf{A}:\mathbf{b}=A_{ijkl}b_{lk}$. In \eqref{sinc} we have neglected higher order terms, i.e.\ those of $O(\boldsymbol{\ga}^2)$. Defining the \textit{push-forward of the incremental nominal stress} as $\boldsymbol{\zeta}=\mathbf{F}\mathbf{s}$, from \eqref{sinc} this therefore has the form
\begin{align}
\boldsymbol{\zeta} &= \mathbf{M}:\boldsymbol{\gamma} +q \mathbf{I} - Q \boldsymbol{\gamma} \label{zetainc}
\end{align}
where the components of the instantaneous modulus tensor $\mathbf{M}$ are defined by
\begin{align}
M_{ijkl} &= \derivtwomix{\mathcal{W}}{F_{jm}}{F_{ln}}F_{im}F_{kn}. \label{Minc}
\end{align}
Convenient forms for this tensor have been given by Ogden (1997) as
\begin{align}
M_{iijj} &= \la_i\la_j \mathcal{W}_{ij} = M_{jjii}, \hspace{1.7cm}\tn{no sum on $i,j$}, \\
M_{ijij} &= \begin{cases}
\left(\frac{\la_i \mathcal{W}_i-\la_j \mathcal{W}_j}{\la_i^2-\la_j^2}\right)\la_i^2, & i\neq j, \la_i\neq \la_j, \\
\la_i^2\mathcal{W}_{ii}-\la_i\la_j \mathcal{W}_{ij}+\la_i \mathcal{W}_i, & i\neq j, \la_i= \la_j,
\end{cases}\\
M_{ijji} &= M_{jiij} = M_{ijij}-\la_i\mathcal{W}_i, \hspace{1cm}i\neq j,
\end{align}
where $\mathcal{W}_i=\pa \mathcal{W}/\pa\la_i$ and $\mathcal{W}_{ij}=\pa^2 \mathcal{W}/\pa\la_i\pa\la_j$. Analysis of the Cauchy stress gives rise to the relationship $\boldsymbol{\tau} = \boldsymbol{\zeta} + \boldsymbol{\gamma}\mathbf{T}$ for an incompressible material.

%%%%%%%%%%%%%%%%%%%%%%%%%%%%%

The equations of motion in body $\bar{b}$ in terms of the Cauchy stress are
\begin{align}
\overline{\Div}\bar{\mathbf{T}} = \bar{\rho}\derivtwo{\bar{\mathbf{U}}}{t} = \rho\derivtwo{\mathbf{u}}{t} \label{A10}
\end{align}
where $\overline{\Div}$ denotes divergence in $\bar{x}$, i.e.\ with respect to the configuration $\bar{b}$. Equation \eqref{A10} has this form since the body is incompressible ($\bar{\rho}=\rho$) and $\bar{\mathbf{U}}=\mathbf{U}+\mathbf{u}$ where $\mathbf{U}$ is independent of time.  Using the fact that $\Div\mathbf{T}=0$, it can be shown that for an incompressible medium
\begin{align}
\overline{\Div}\bar{\mathbf{T}} = \Div(\boldsymbol{\tau}-\boldsymbol{\ga}\mathbf{T}) = \Div\boldsymbol{\zeta}
\end{align}
where we have used the relationship between $\boldsymbol{\zeta}$ and $\boldsymbol{\tau}$ above. The incremental equations are therefore
\begin{align}
\Div\boldsymbol{\zeta} = \rho\derivtwo{\mathbf{u}}{t} \label{A12}
\end{align}
together with appropriate forcing terms as required. Indeed, time-harmonic line source forcing in the $z$ direction gives rise to incremental horizontally polarized shear waves of the form \eqref{uwSH}. Two of the equations \eqref{A12} imply that $q=0$ and the remaining single governing equation of motion is
\begin{align}
\frac{1}{r}\deriv{}{r}\left(r\zeta_{rz}\right)+\frac{1}{r}\deriv{\zeta_{\theta z}}{\theta}+\rho\om^2 w = \frac{C}{r_0}\de(r-r_0)\de(\theta-\theta_0). \label{eomzetaapp}
\end{align}
where
\begin{align}
\zeta_{rz} &= M_{rzrz}\gamma_{zr}, & \zeta_{\theta z} &= M_{\theta z \theta z}\gamma_{z\theta}. \label{eomzetarelapp}
\end{align}
In this case $\tau_{rz} = \zeta_{rz}$ and $\tau_{\theta z} = \zeta_{\theta z}$ and so the instantaneous moduli $M_{rzrz}$ and $M_{\theta z \theta z}$ can be considered as pressure dependent anisotropic shear moduli which for an incompressible neo-Hookean material are
\begin{align}
\mu_r= M_{rzrz} &= \mu\la_r^2 = \frac{\mu}{L}\left(\frac{r^2+M}{r^2}\right), &
\mu_{\theta}=M_{\theta z\theta z}  &= \mu\la_{\theta}^2 = \frac{\mu}{L}\left(\frac{r^2}{r^2+M}\right).
\label{murmut}
\end{align}
Using \eqref{eomzetarelapp} with \eqref{murmut} in \eqref{eomzetaapp} gives the governing equation of motion in terms of the displacement $w$ which is
\begin{align}
\frac{1}{r}\deriv{}{r}\left(r\mu_r(r)\deriv{w}{r}\right)+\frac{\mu_{\theta}(r)}{r^2}\derivtwo{w}{\theta}+\rho\om^2 w &= \frac{C}{r_0}\de(r-r_0)\de(\theta-\theta_0). \label{incrapp}
\end{align}

\section{Scattering of an antiplane elastic line source by a cylindrical cavity} \label{unstressed}

%In this section, in order to define notation and for later reference, we summarize results regarding the scattering of time-harmonic horizontally-polarized shear (SH) waves from a cylindrical cavity due to an incident field generated by a line source (cf.\ the case of acoustic scattering (Parnell \& Abrahams 2010)).

Consider an isolated cylindrical cavity of radius $A$, inside an isotropic homogeneous elastic host region of infinite extent in all directions. We specify a Cartesian coordinate system $(X,Y,Z)$, with origin at the centre of the cavity and whose $Z$ axis runs parallel with the axis of the cylindrical cavity. Consider the problem of scattering by a time-harmonic line source (of small-amplitude) which is polarized in the $Z$ direction and located at $X=X_0, Y=Y_0$ (a distance $R_0$ away from the centre of the cylinder and at an angle $\Theta_0\in[0,2\pi)$ subtended from the $X$-axis). This forcing therefore gives rise to linear elastic waves polarized in the $Z$ direction propagating in the $XY$ plane, i.e.\ SH waves. The displacement field is therefore given by $\textbf{U}=\Re\left[W(X,Y)\mathbf{e}_Z\exp(-i\om t)\right]$, where $\mathbf{e}_Z$ is a unit basis vector in the $Z$ direction and $\om$ is the circular frequency and where $W(X,Y)$ is governed by (see e.g.\ Graff (1975))
\begin{align}
(\mu\nabla^2 + \rho\om^2)W &= C\de(X-X_0)\de(Y-Y_0) = \frac{C}{R_0}\de(R-R_0)\de(\Theta-\Theta_0). \label{inhomprob0}
\end{align}
Here we have defined the planar cylindrical polar coordinates $R, \Theta$ via the standard relations $X=R\cos\Theta, Y=R\cos\Theta$ and we note that $R_0$ and $\Theta_0$ are thus defined by $X_0=R_0\cos\Theta_0, Y_0=R_0\cos\Theta_0$. We also point out that $C$ is the force per unit length (in the $Z$ direction) of the imposed line source and $\rho$ and $\mu$ are the mass density and shear modulus of the host medium. We re-write \eqref{inhomprob0} in the form
\begin{align}
(\nabla^2 + K^2)W &= \frac{C}{\mu R_0}\de(R-R_0)\de(\Theta-\Theta_0), \label{inhomprob}
\end{align}
where the wavenumber $K$ is defined by $K^2=\rho\om^2/\mu$. The traction-free boundary condition on the surface of the cavity at $R=A$ is
\begin{align}
\mu\deriv{W}{R} &= 0. \label{BC111}
\end{align}
We can write the solution of this problem in the form
\begin{align}
W &= W_i + W_s \label{totfield}
\end{align}
where $W_i$ and $W_s$ represent incident and scattered (outgoing) fields from the cylindrical cavity. In particular, the incident field represents the solution of the inhomogeneous problem \eqref{inhomprob} which is outgoing from the source location, i.e.
\begin{align}
W_i &= \frac{C}{4i\mu}\tn{H}_0(KS) \label{w0scat}
\end{align}
where $S=\sqrt{(X-X_0)^2+(Y-Y_0)^2}$. We have defined $\tn{H}_0(KS)=\tn{H}_0^{(1)}(KS)=\tn{J}_0(KS)+i\tn{Y}_0(KS)$, the Hankel function of the first kind, noting that $\tn{J}_0$ and $\tn{Y}_0$ are Bessel functions of the first and second kind respectively, of order zero. Together with the time dependence in the problem, this ensures an outgoing field from the source. Graf's addition theorem allows us to write this field relative to the coordinate system $(R,\Theta)$ centred at the origin of the cavity in the form (Martin 2006)
\begin{align}
W_i &= \frac{C}{4i\mu}\tn{H}_0(KS) = \frac{C}{4i\mu}\times\begin{cases}
\sum_{n=-\infty}^{\infty} \tn{H}_n(KR_0)\tn{J}_n(KR)e^{in(\Theta-\Theta_0)}, & R<R_0,\\
\sum_{n=-\infty}^{\infty} \tn{H}_n(KR)\tn{J}_n(KR_0)e^{in(\Theta-\Theta_0)}, & R>R_0,
\end{cases}
\label{H0scat}
\end{align}
where $\tn{H}_n$ and $\tn{J}_n$ are respectively Hankel and Bessel functions of the first kind, and of order $n$. Upon writing the scattered field in the form
\begin{align}
W_s &= \sum_{n=-\infty}^{\infty} (-i)^n A_n\tn{H}_n(KR)e^{in(\Theta-\Theta_0)} \label{Wsfld}
\end{align}
and using this together with \eqref{totfield} and the first of \eqref{H0scat} in \eqref{BC111} we find that the scattering coefficients due to the incident wave from the line source are
\begin{align}
A_n &= \frac{C(-1)^n}{4\mu i^{n-1}}\frac{\tn{J}_n'(KA)}{\tn{H}_n'(KA)}\tn{H}_n(KR_0). \label{An1}
\end{align}
Let us now consider the limit as $R_0\rightarrow \infty$ in \eqref{w0scat} and the first of \eqref{H0scat}. Using the fact that (p.\ 364 of Abramowitz \& Stegun (1965))
\begin{align}
H_n(KR_0) \sim (-i)^n\sqrt{\frac{2}{\pi K R_0}}e^{i(KR_0-\pi/4)}
\end{align}
as $R_0\rightarrow\infty$ and upon setting
\begin{align}
C &= 2i\mu\sqrt{2\pi KR_0}e^{i(\pi/4-KR_0)}, \label{Cconst}
\end{align}
we find that in this limit,
\begin{align}
W_i &\sim \sum_{n=-\infty}^{\infty}i^n \tn{J}_n(KR)e^{in(\Theta-\theta_{inc})} = e^{iK(X\cos\theta_{inc}+Y\sin\theta_{inc})},
\end{align}
which is an incident plane wave, propagating at an angle of incidence $\theta_{inc}=\Theta_0-\pi\in[-\pi,\pi)$ to the $X$ axis (since $\Theta_0\in[0,2\pi)$). The scattering coefficients appearing in \eqref{Wsfld} for this forcing are thus
\begin{align}
A_n^{(pw)} &= -\frac{\tn{J}_n'(KA)}{\tn{H}_n'(KA)}, \label{An2}
\end{align}
where the superscript $(pw)$ indicates a plane-wave forcing. These are the standard scattering coefficients associated with plane wave scattering from a circular cylindrical cavity (see e.g.\ Martin (2006), p.\ 123, equation (4.5)), noting the form of the scattered field in \eqref{Wsfld} with $\Theta_0=\theta_{inc}+\pi$.

%Considering the incident plane wave case, taking the limit as $KA\rightarrow 0$ in \eqref{An2} we find that the leading order coefficients are
%\begin{align}
%A_0^{(pw)} &= \frac{\pi}{4i}(KA)^2,  &  A_{-1}^{(pw)} = A_1^{(pw)} &= -\frac{\pi}{4i}(KA)^2 \label{A0A1}
%\end{align}
%and all higher order terms are of $o((KA)^2)$. Furthermore, one can show using classical multiple scattering theory with the Lax Quasi-Crystalline Approximation, in the low frequency limit, the effective anti-plane shear modulus of a fibre reinforced  composite can be written as (Bose \& Mal 1973, Parnell \& Abrahams 2010)
%\begin{align}
%\frac{\mu_*}{\mu_0} &= \frac{1+4i A_1/(K^2|\mathcal{D}|)}{1-4i A_1/(K^2|\mathcal{D}|)}
%\end{align}
%where $A_1$ is the scattering coefficient of the fibre given plane wave incidence and where $|\mathcal{D}|$ is the total planar area of the fibre-reinforced material in question. Therefore for a distribution of cavities, using \eqref{A0A1} we find
%\begin{align}
%\frac{\mu_*}{\mu_0} &= \frac{1 - \phi}{1+\phi}
%\end{align}
%where $\phi$ is the volume fraction (per unit length in the $Z$ direction) of cavities. We shall assess how this effective property becomes modified due to nonlinear elastic pre-stress.

\subsection*{Acknowledgements}
The author acknowledges useful discussions with Prof.\ I.\ David Abrahams (University of Manchester) relating to this work. He is also grateful to Michael Croucher (University of Manchester) for assistance with figures 1 and 4.

%%%% Bibliography  %%%%%%%%%%

%\label{lastpage}

\end{document}